\documentclass[aps,pre,twocolumn,groupedaddress,showpacs,floatfix]{revtex4}
\usepackage{amsmath}
\usepackage{amssymb}
\usepackage{graphicx}

\begin{document}

\title{An Analytical Approach to Neuronal Connectivity} 

\author{Luciano da Fontoura Costa} 
\affiliation{Institute of Physics of S\~ao Carlos. 
University of S\~ ao Paulo, S\~{a}o Carlos,
SP, PO Box 369, 13560-970, 
phone +55 162 73 9858,FAX +55 162 71
3616, Brazil, luciano@if.sc.usp.br}

\date{\today}

\begin{abstract}   

This paper describes how realistic neuromorphic networks can have
their connectivity properties fully characterized in analytical
fashion.  By assuming that all neurons have the same shape and are
regularly distributed along the two-dimensional orthogonal lattice
with parameter $\Delta$, it is possible to obtain the accurate number
of connections and cycles of any length from the autoconvolution
function as well as from the respective spectral density derived from
the adjacency matrix.  It is shown that neuronal shape plays an
important role in defining the spatial spread of network connections.
In addition, most such networks are characterized by the interesting
phenomenon where the connections are progressively shifted along the
spatial domain where the network is embedded.  It is also shown that
the number of cycles follows a power law with their respective length.
Morphological measurements for characterization of the spatial
distribution of connections, including the adjacency matrix spectral
density and the lacunarity of the connections, are suggested.  The
potential of the proposed approach is illustrated with respect to
digital images of real neuronal cells.

\end{abstract}


\maketitle

A particularly meaningful way to understand neurons is as cells
optimized for \emph{selective connections}, i.e. connecting between
themselves in a specific manner so as to achieve proper circuitry and
behavior.  Indeed, the intricate shape of dendritic trees provide the
means for connecting with specific targets while minimizing both the
cell volume and the implied metabolism (e.g. \cite{Purves:1985,
Karbowski:2001}).  While great attention has been placed on the
importance of synaptic strength over the emerging neuronal behavior,
geometrical features such as the shape and spatial distribution of the
involved neurons play a particularly important role in defining the
network connectivity.  In addition, the topographical organization and
connections pervading the mammals' cortex provide further indication
that adjacencies and spatial relationships are fundamental for
information processing by biological neuronal networks.  The
importance of neuronal geometry has been reflected by the growing
number of related works (see, for instance, \cite{SI_BM:2003}).
However, most of such approaches target the characterization of
neuronal morphology in terms of indirect and incomplete measures such
as area, perimeter and fractal dimension of the dendritic and axonal
arborizations, to name but a few.  Interesting experimental results
regarding the connectivity of neuronal cells growth \emph{in vitro}
have been reported in \cite{Shefi:2002, Segev:2003} and what is
possibly the first direct computational approach to neuronal
connectivity was only recently reported in \cite{Percolation:2003},
involving the experimental estimation of the critical percolation
density as neuronal cells are progressively superposed onto a
two-dimensional domain.  At the same time, the recent advances in
complex network formalism (e.g. \cite{Albert_Barab:2002,
Bollobas:2002, Barabasi_Ravasz:1998, Buckley:1990, Amaral:2000})
provide a wealthy of concepts and tools for addressing connectivity.
Initial applications of such a theory to bridge the gap between
neuronal shape and function were reported in
\cite{Stauffer:2003,Costa_BM:2003}.

As such developments are characterized by computational approaches
involving numerical methods and simulation, a need arises to develop
an analytical framework for neuromorphic characterization that could
lead to additional insights and theoretical results regarding the
relationship between neuronal shape and function.  The current paper
describes how regular neuromorphic complex networks can be obtained
and their connectivity fully characterized in analytical terms.  By
regular it is meant that all cells have the same shape and are
regularly distributed along the two-dimensional orthogonal lattice
with parameter $\Delta$.  Such a kind of networks can be considered as
models of biological neuronal networks characterized by planarity and
morphologic regularity, as is the case with ganglion cell retinal
mosaics \cite{Wassle:1974} and the basal dendritic arborization of
cortical pyramidal cells.

\begin{figure}
 \begin{center} 
   \includegraphics[scale=.4,angle=-90]{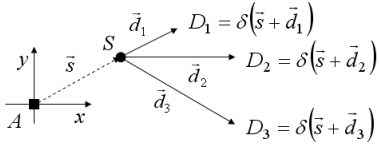}
   \caption{The geometry of a simplified neuronal cells represented in
 terms of its axon $A$, soma centroid $S$ and dendrite points
 $D_i$.~\label{fig:neurmod}} \end{center}
\end{figure}

Let the neuronal cell be represented in terms of the triple
$\eta=[A,S,D]$ where $A$ is the set of points belonging to its axonal
arborization, $S$ is the set of points corresponding to the respective
soma (neuronal body) and $D$ are the dendritic arborization points.
For simplicity's sake, a finite and discrete neuronal model is
considered prior to its continuous general formulation.  We therefore
assume that the points used to represent the neuron belong to the
square orthogonal lattice $\Omega= \left\{ 1, 2, \ldots, N \right\}
\times \left\{ 1, 2, \ldots, N \right\}$, with parameter $\Delta =1$.
The axon and soma are represented by a single point each,
i.e. $A={\vec{a}}$ and $S={\vec{s}}$.  Such points could be understood
as corresponding to the tip of the axon and the soma center of mass,
respectively.  The dendritic arborization is represented in terms of
the finite set of dendrite points $D={D_1, D_2, \ldots, D_M}$, and it
is henceforth assumed that a dendrite point never coincides with the
axon.  Figure~\ref{fig:neurmod} illustrates such a geometrical
representation for a neuron with 3 dendrite points.  Observe that the
coordinate origin coincides with the axon, which is taken as reference
for the soma and dendrite coordinates.

\begin{figure}
 \begin{center} \includegraphics[scale=.4,angle=-90]{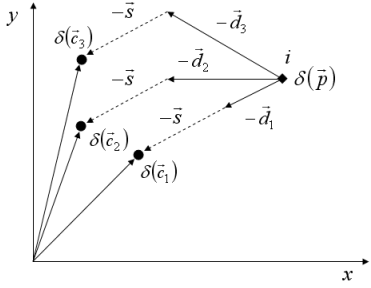}
   \caption{The neurons connected to neuron $i$ through unit-length
   paths can be obtained through the convolution between the position
   of neuron $i$ with the dendritic shape.~\label{fig:constr}}
 \end{center}
\end{figure}

Neuromorphic networks (actually digraphs \cite{Bollobas:2002}) can now
be obtained by placing one such a neuron at all possible nodes of the
orthogonal lattice $\Omega$.  By fully characterizing such neuronal
architectures, the connections established whenever an axon is
overlaid onto a dendrite point (no connections with soma are allowed)
stand out as particularly important features of the obtained network.
Consequently, it is important to obtain analytical expressions fully
characterizing neuronal connectivity, in the sense of the spatial
distribution of paths and cycles of any specific length.  We start by
considering the connections initiated from a single specific neuron
$i$ placed at position $\vec{p}$.  As illustrated in
Figure~\ref{fig:constr}, the three neurons identified by the vectors
$\vec{c}_1=\vec{p}-\vec{d}_1-\vec{s}$,
$\vec{c}_2=\vec{p}-\vec{d}_2-\vec{s}$,
$\vec{c}_3=\vec{p}-\vec{d}_3-\vec{s}$ are directly connected to $i$
through paths of unit length.  As is clear from such a construction,
the set of neurons connected to $i$ through unit-length paths can be
obtained by convolving the initial point $\delta \left\{\vec{p}
\right\}$ with the function $g(x,y)=\delta \left\{-\vec{d}_1-\vec{s}
\right\} + \delta \left\{-\vec{d}_2-\vec{s} \right\} + \delta
\left\{-\vec{d}_3-\vec{s} \right\}$.  In other words, given a set of
initial neurons with axons located at $\xi(x,y)$, the number
$\chi(x,y)$ of connections received from that set by a neuron at
position $(x,y)$ is obtained as in Equation~\ref{eq:chi}.  The
binary-valued (1-true, 0-false) function $\nu(x,y)$ given in
Equation~\ref{eq:nu} indicates whether there is a unit-length path
between the neuron at $(x,y)$ and the initial set of points
$\xi(x,y)$, where $\phi()$ is the hard-limiting function.  The
functions expressing the number of connections of length $k$ between
$\xi(x,y)$ and the neuron at position $(x,y)$ and the presence of at
least one such a connection at that position are given by
Equations~\ref{eq:chik} and~\ref{eq:nuk}, respectively.  The total
number of connections from length 1 to $k$ received by the neuron
located at position $(x,y)$ from $\xi(x,y)$ is given as in
Equation~\ref{eq:tau}.  Observe that the use of the Dirac delta
function in such a formulation allows the immediate extension of such
results to continuous spatial domains.  It should be also observed
that the above framework can be immediately extended to generalized
axons by having the dendritic arborization to undergo Minkowski
dilation \cite{Stoyan:1995} with the axonal shape and considering as
axon only the single point corresponding axon a reference along its
shape.  While the analytical characterization of the connectivity of
the considered network models has been allowed by the fact that
identical neuronal shapes are distributed along all points of the
orthogonal lattice, it is interesting to consider extensions of such
an approach to other situations.  An immediate possibility is to
consider sparser configurations, characterized by larger lattice
parameters $\Delta$.  Such an extension involves sampling the neuronal
cell image at larger steps.

\begin{eqnarray}
  \chi(x,y)=g(x,y) \ast \xi(x,y) \label{eq:chi} \\
  \nu(x,y)=\phi(\chi(x,y)) \label{eq:nu} \\ \chi_k(x,y)=\underbrace{
  g(x,y) \ast \ldots \ast g(x,y) }_{k \times} \xi(x,y) \label{eq:chik}
  \\ \nu_k(x,y)=\phi(\chi_k(x,y)) \label{eq:nuk} \\
  \tau_k(x,y)=\sum_{j=1}^{k}(\chi_j(x,y)) \label{eq:tau}
\end{eqnarray}

\begin{figure}
 \begin{center} \vspace{0.3cm}
  \includegraphics[angle=-90,scale=0.5]{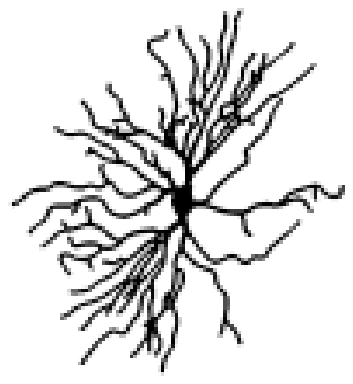} \hspace{1.5cm}
  \includegraphics[angle=-90,scale=0.5]{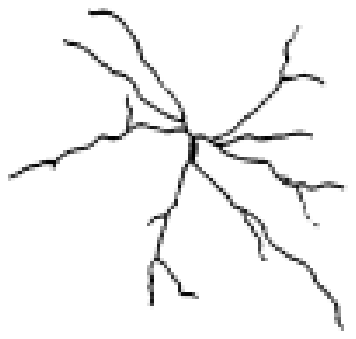} \\
  \vspace{0.2cm} (a) \hspace{2.5cm} (b) \\
  \includegraphics[angle=0,scale=0.4]{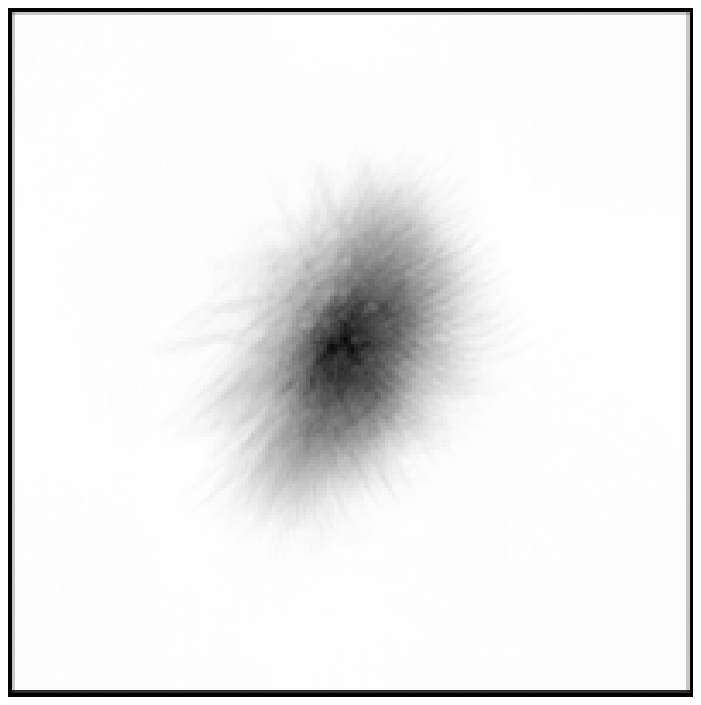} \hspace{0.5cm}
  \includegraphics[angle=0,scale=0.4]{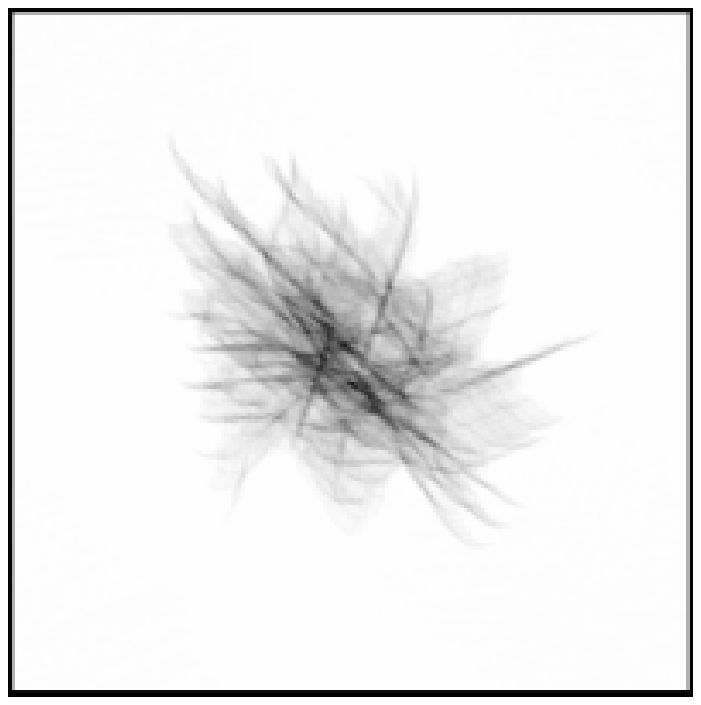} \\
  \vspace{0.2cm} (c) \hspace{2.5cm} (d) \\
  \includegraphics[angle=0,scale=0.4]{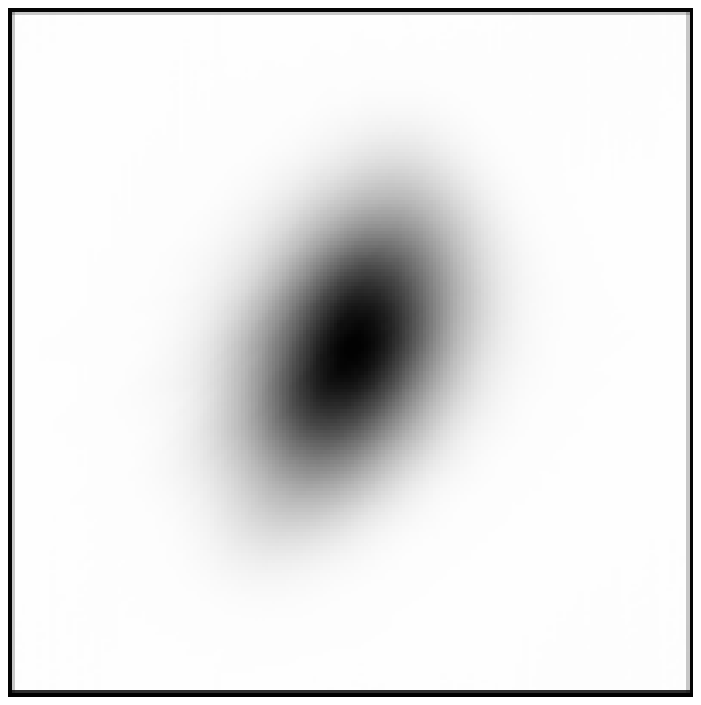} \hspace{0.5cm}
  \includegraphics[angle=0,scale=0.4]{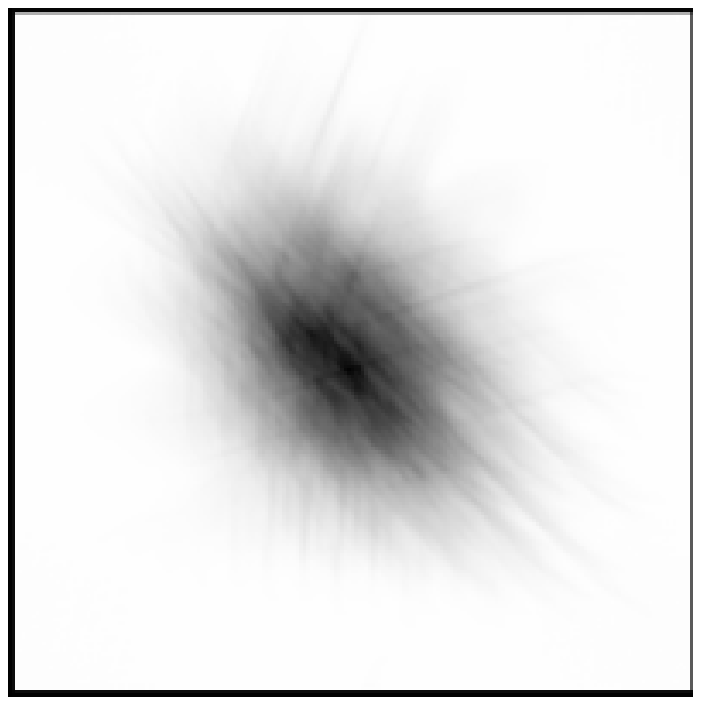} \\
  \vspace{0.2cm} (e) \hspace{2.5cm} (f) 

  \caption{Two real neuronal cells (a-b) and their respective total
  number of connections of length $k=1$ (c-d) and 2 (e-f).  The axon
  has been placed at the cell centroid (considering soma plus
  dendrites).  The neuronal cell figures in (a) and (b) are adapted
  with permission from \cite{Wassle:1974}.~\label{fig:ex1}} \end{center}
\end{figure}

\begin{figure}
 \begin{center} \vspace{0.3cm}
  \includegraphics[angle=-90,scale=0.4]{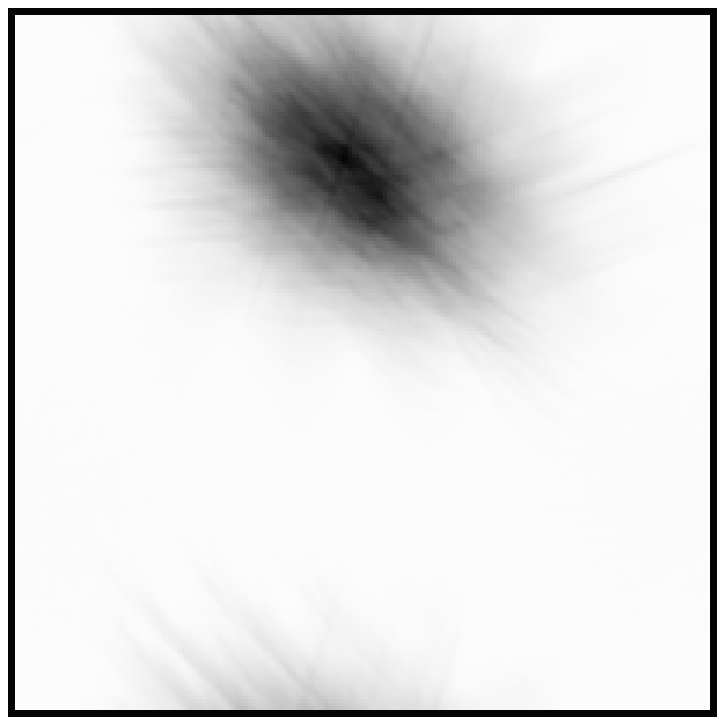} \hspace{0.1cm}
  \includegraphics[angle=-90,scale=0.4]{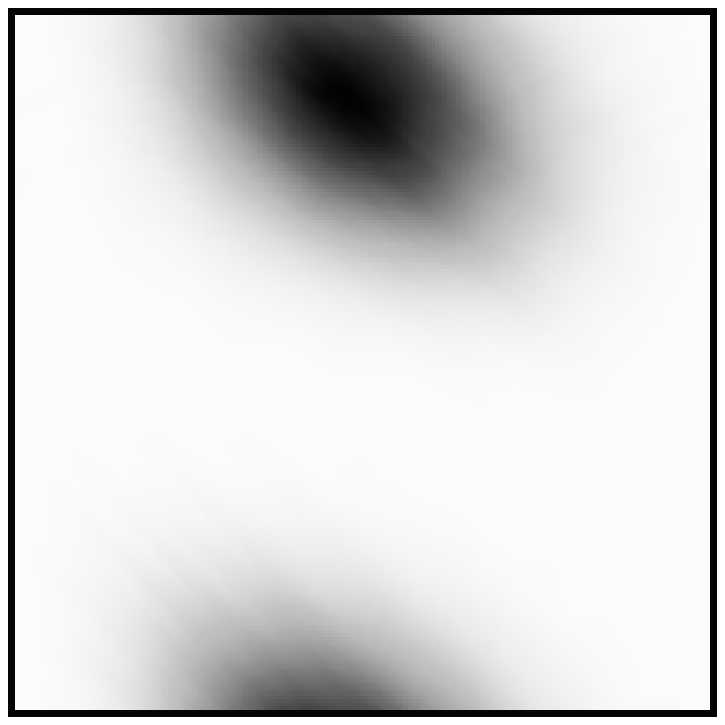} \\
  \vspace{0.5cm} (a) \hspace{2.5cm} (b) \\
  \includegraphics[angle=-90,scale=0.4]{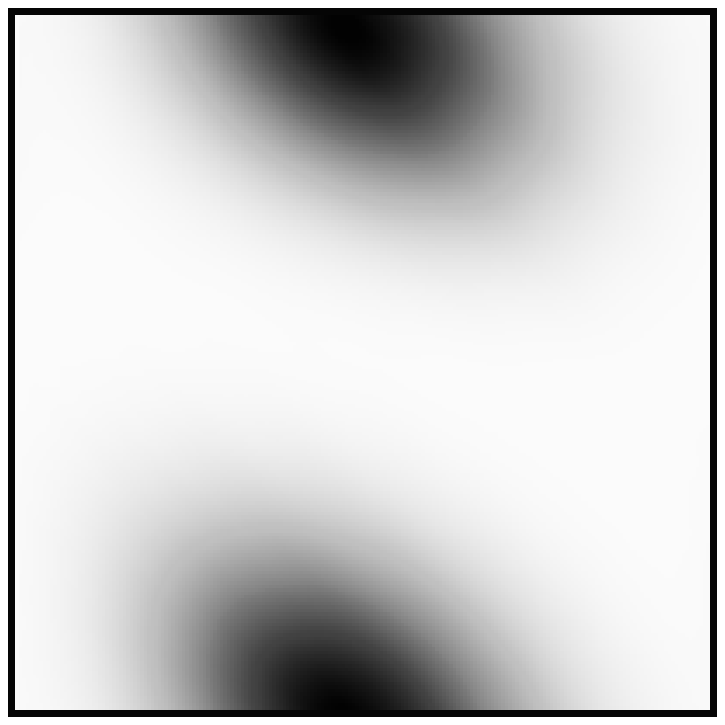} \hspace{0.1cm}
  \includegraphics[angle=-90,scale=0.4]{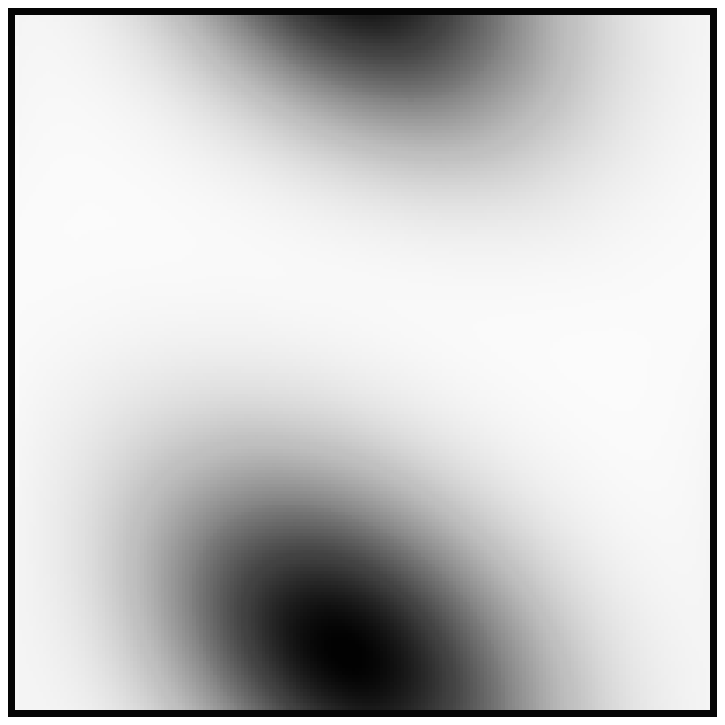} \\
  \vspace{0.5cm} (c) \hspace{2.5cm} (d) 

  \caption{The total number of connections of length $k=1$ (a), 2 (b),
  3 (c) and 4(d) for the neuronal cell in Figure~\ref{fig:ex1}(a) with
  the axon placed over the cell centroid, which is itself displaced
  from the cell centroid by $ \vec{s}=(0,7)$ pixels.~\label{fig:ex2}}
  \end{center}
\end{figure}

Figure~\ref{fig:ex1} shows two digital images obtained from real
ganglion cells, (a) and (b), and their respective functions
$\chi_k(x,y)$ for $k=2$ to $4$.  The axon has been placed over the
centroid of the neuronal shape (including soma and dendrites), whereas
the dendritic trees have been spatially sampled into 2033 and 671
pixels, respectively.  Figure~\ref{fig:ex2} shows $\chi_k(x,y)$
obtained for the cell in Figure~\ref{fig:ex1}(a) but with the soma
located at the cell center of mass, which is displaced from the cell
centroid by $\vec{s}=(0,7)$.  It is clear from such results that the
neuron morphology strongly determines the connectivity between cells
in two important senses: (i) the spatial scattering of the dendrite
points influences the spatial connectivity distribution and (ii) the
relative position of the axon defines how the centroid of the
connections shifts for increasing values of $k$.  While the increased
number of synaptic connections implied by denser neuronal shapes is as
expected, it is clear from the example in Figure~\ref{fig:ex2} that
the distance from axon to cell center of mass implies a coordinated
spatial shifting of connections as their respective lengths are
increased.  Given the predominantly two-dimensional structure of the
mammals' cortex, such an effect provides an interesting means to
transmit information horizontally along such structures.  As the
proper characterization, classification, analysis and simulation of
neuromorphic networks are all affected by these two interesting
phenomena, it is important to derive objective related measurements.
Let $P(x,y)$ be a density function obtained by normalizing
$\chi_k(x,y)$.  Thus, the spatial scattering of the connections can be
quantified in terms of the respective covariance matrix $K_k$, and the
spatial displacement of the centroid of $P(x,y)$ can be quantified in
terms of the `speed' $v=|| \vec{s} ||$.  Additional geometrical
measurements of the evolution of the neuronal connectivity that can be
derived from the covariance matrix $K$ include the angle $\alpha_k$
that the distribution main axis makes with the x-axis and the ratio
$\rho_k$ between the largest and smallest respective eigenvalues.

Another interesting network feature related to connectivity is its
number $C_{\ell,k}$ of cycles of length $\ell$ established by the
synaptic connections.  This feature can be calculated from the
enlarged matrix $A$ obtained by stacking the columns of the matrix
where the neuronal cell image is represented in order to obtain the
rows of $A$, while the reference point of the cell is shifted along
the main diagonal of $A$. Observe that the image size $N \times N$ has
to be large enough in case dynamics near the toroidal boundary
conditions are to be avoided.  The $N^2$ eigenvalues of the thus
obtained \emph{adjacency matrix} \cite{Albert_Barab:2002} of the whole
two-dimensional network are henceforth represented as $\lambda_i$, $i
= 1, 2, \ldots, N^2$.  As $A$ is circulant, these eigenvalues can be
immediately obtained from the Fourier transform of its first row.
Observe that the simplicity and speed of such an approach allow for
systematic investigation of a variety of different neuronal shapes.
As the cell reference point is assumed never to coincide with a
dendrite point, we also have that $\sum_{r=1}^{N} \lambda_{r} = 0$. As
$A$ is a non-negative matrix, there will always be a non-negative
eigenvalue $\lambda_M$, called the \emph{dominant eigenvalue of A},
such that $\lambda_{r} \leq \lambda_M$ for any $r=1, 2, \ldots,N$.
The (unnormalized) \emph{cycle density} of $\Gamma$, defined in
Equation~\ref{eq:cyc_distr}, provides a clear characterization of the
network cycles in terms of their respective populations.  The total
number of cycles up to length $P$ is defined as $T=\sum_{p=1}^{P}
\Lambda(p)$.  The \emph{spectral density}
(e.g. \cite{Albert_Barab:2002}) of the adjacency matrix, defined in
Equation~\ref{eq:spec_dens}, where $\lambda_p$ is the $p-$th
eigenvalue of $A$, provides an additional way to characterize the
topology of the obtained networks.

\begin{eqnarray}
  \Phi(t)=\sum_{p=1}^{P} \Lambda(p) \delta(t-p)  \label{eq:cyc_distr} \\
  \rho(\lambda)=\frac{1}{N} \sum_{r=1}^{N} \delta(\lambda-\lambda_r)  
     \label{eq:spec_dens}
\end{eqnarray}

\begin{figure}
 \begin{center} 
   \includegraphics[scale=.43,angle=-90]{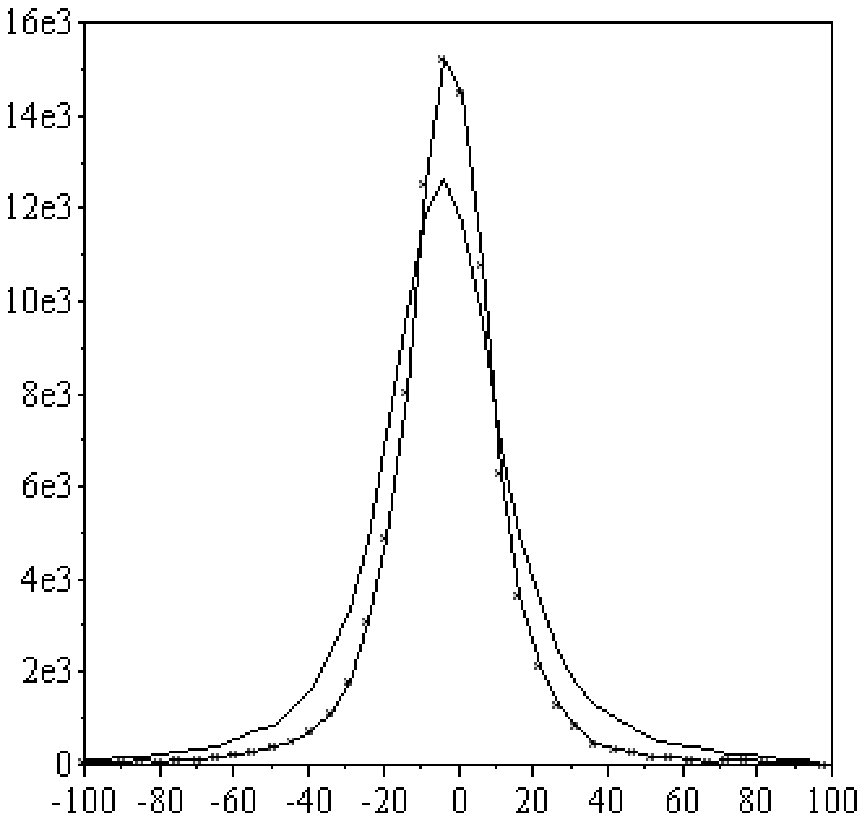} \hspace{0.2cm}
   \includegraphics[scale=.43,angle=-90]{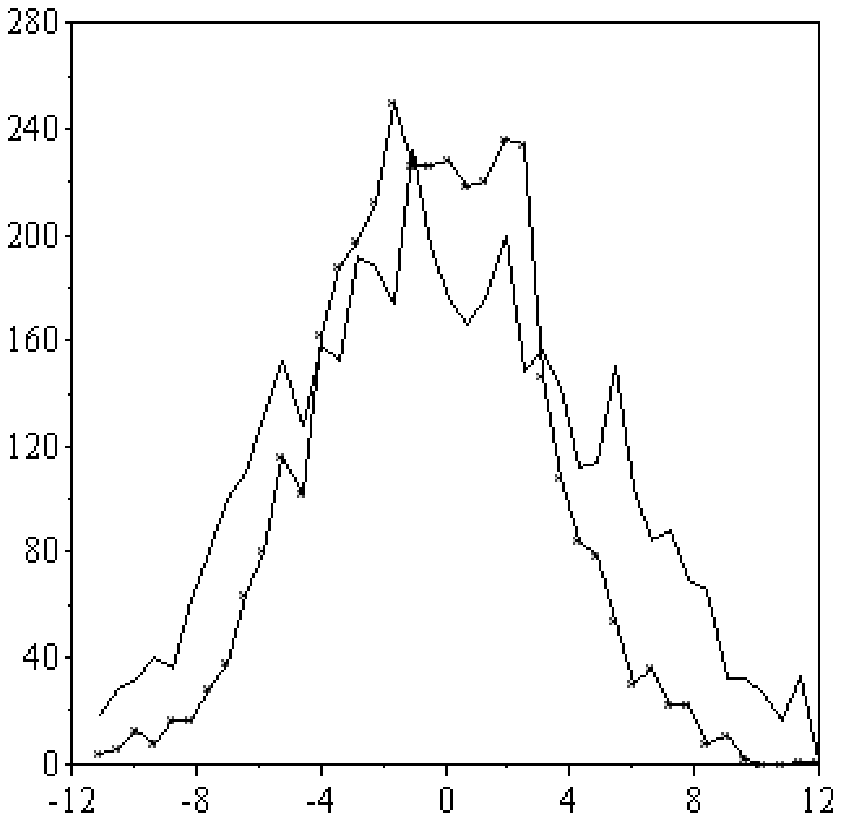} \\
   \vspace{0.2cm} (a) \hspace{4cm} (b) 

   \caption{Spectral density of the adjacency matrices obtained for
   the neuronal cells in Figure~\ref{fig:ex1} considering $\Delta=1$
   (a) and $\Delta=5$ (b).  The crossed lines referes to the
   sparser neuronal cell.~\label{fig:eigs}} \end{center}
\end{figure}

It is interesting to note that as the maximum length $p$ in
Equation~\ref{eq:cyc_distr} increases, the number of cycles with that
length can be approximated as $\Lambda(p) \cong \lambda_M^p$, i.e. the
dynamics of $p$ is defined by the dominant eigenvalue of $A$, and the
distribution of $\Lambda(p)$ tends to follow a power law.  The
eigenvalue $\lambda_M$, which depends on the specific dynamics through
which new edges are incorporated into the network, therefore
represents an interesting parameter for characterizing the cyclic
composition of complex networks.  Figures~\ref{fig:eigs}(a) and (b)
show the real part (recall that the adjacency matrix for a digraph is
not necessarily symmetric) of the spectral density of the adjacency
matrices obtained for the neuronal cells in Figure~\ref{fig:ex1}(a)
and (b) considering $\Delta=1$ and 5.  The wider dispersion of the
spectrum of the denser cell in Figure~\ref{fig:eigs}(a) reflects a 
higher potential for connections of that neuron in both cases.  It is
also clear that the separation of cells by $\Delta=5$ leads to a 
substantially smaller spectrum, with immediate implications for the
respective neuronal connectivity.

\begin{figure}
 \begin{center} 
   \includegraphics[scale=.42,angle=-90]{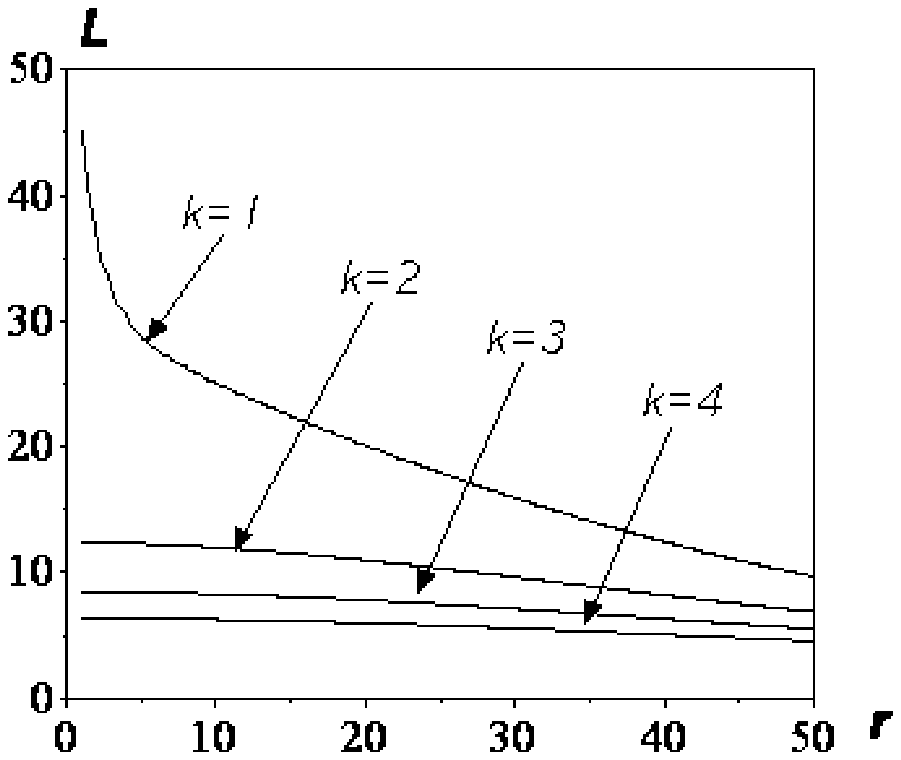} 
   \includegraphics[scale=.42,angle=-90]{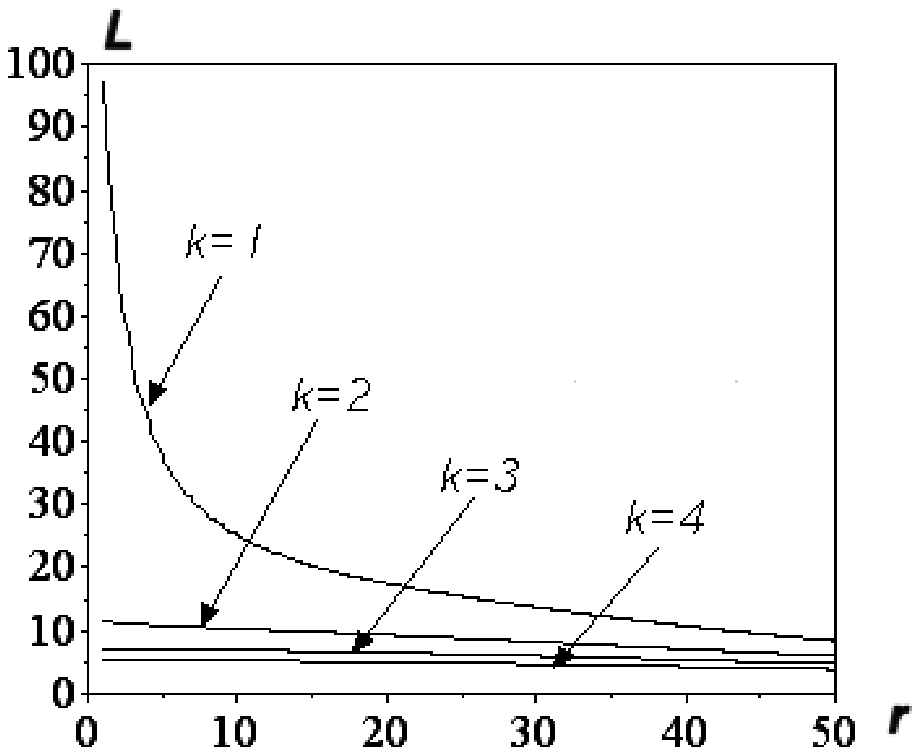} \\
   \vspace{0.2cm} (a) \hspace{4cm} (b) 

   \caption{Lacunarities for the spatial distributions of connections 
   for the cells in Figure~\ref{fig:ex1}.~\label{fig:lacuns}} \end{center}
\end{figure}

An additional morphological property of the spatial distribution of
the connections is their respective \emph{lacunarity}
(e.g. \cite{Hovi:2003}), which expresses the degree of translational
invariance of the obtained densities.  Figure~\ref{fig:lacuns} shows
the lacunarities of the connection densities obtained for the two
considered cells with respect to $k=1$ to $4$.  It is interesting to
observe that most of the lacunarity differences are observed for
$k=1$, with similar curves being obtained for larger values of $k$.
At the same time, the denser cell led to lower lacunarity values.
Given their immediate implications for neuronal connectivity, the
above proposed set of neuronal shape measurements present specially
good potential for neuron characterization and classification.

In addition to paving the way for the complete analytical
characterization of the connections in regular neuromorphic networks
in terms of paths and cycles, the framework proposed in this article
can be immediately adapted to express the spread of neuronal activity
starting from the stimulus $\xi(x,y)$.  The neurons are understood to
accept input and produce output in synchronous manner at each clock
cycle $T$.  For instance, the situation where the neuronal cell output
corresponds to the inner product between its shape and the respective
area of the input space can be immediately characterized in terms of
the eigenvalues and eigenvectors of precisely the same adjacency
matrix $A$ constructed as described above.  Although the proposed
methodology assumes identical, uniformly distributed neuronal cells,
it is expected that they provide a reference model for investigating
and characterizing real networks characterized by a certain degree of
regularity, such as some subsystems found in the retina and cortex.
Mean-field extensions of the reported approach are currently being
investigated.

\begin{acknowledgments}
The author is grateful to FAPESP (processes 99/12765-2 and 96/05497-3)
and CNPq for financial support.
\end{acknowledgments}

 
\bibliography{graphm}

\end{document}